\documentclass[prl,twocolumn,aps,showpacs,amsmath,amssymb,floatfix]{revtex4-1}

\usepackage{color}
\usepackage{bm}
\usepackage{MnSymbol}
\usepackage{graphicx}
\usepackage{braket}

\renewcommand{\v}[1]{\textbf{#1}}

\usepackage{bm}
\usepackage{graphicx}
\usepackage{amssymb}
\usepackage{color}
\usepackage{empheq}

\newcommand{\tr}{\text{tr}}

\begin{document}

\title{Quantum field theory of nematic transitions in spin orbit coupled spin-1 polar bosons}

\author{E. J. K\"onig}
\author{J. H. Pixley}
\affiliation{Department of Physics and Astronomy, Center for Materials Theory, Rutgers University, Piscataway, NJ 08854 USA}

\date{\today}

\begin{abstract}
We theoretically study an ultra-cold gas of spin-1 polar bosons in a one dimensional continuum which are subject to linear and quadratic Zeeman fields and a Raman induced spin-orbit coupling. 
Concentrating on the regime in which
the background fields can be treated perturbatively 
we analytically solve the model in its low-energy sector, i.e. we characterize the relevant phases and the quantum phase transitions between them. Depending on the sign of the effective quadratic Zeeman field $\epsilon$, two superfluid phases with distinct nematic order appear. In addition, we uncover a spin-disordered superfluid phase at strong coupling. We employ a combination of renormalization group calculations and duality transformations
to access the nature of the phase transitions. At $\epsilon = 0$, a line of spin-charge separated pairs of Luttinger liquids divides the two nematic phases and the transition to the spin disordered state at strong coupling is of the Berezinskii-Kosterlitz-Thouless type. In contrast, at $\epsilon \neq 0$, the quantum critical theory separating nematic and strong coupling spin disordered phases contains a Luttinger liquid in the charge sector that is coupled to a Majorana fermion in the spin sector (i.e. the critical theory at finite $\epsilon$ maps to a quantum critical Ising model that is coupled to the charge Luttinger liquid). Due to an emergent Lorentz symmetry, both have the same, logarithmically diverging velocity. We discuss the experimental signatures of our findings that are relevant to ongoing experiments in ultra-cold atomic gases of $^{23}$Na.
\end{abstract}
\date{\today}

\maketitle

The interplay of internal quantum states and strong interactions can lead to the emergence of new quantum phases of matter and criticality. 
For example, while spin-1/2 quantum magnets can only sustain conventional magnetic order, larger spin systems allow for order in higher angular momentum channels involving 
multipole moments in large spin systems \cite{IkedaMatsuda2012,KoitzschInosov2016,MartelliPaschen2017}. Spinful ultra-cold atomic gases are a particularly fruitful setting to study magnetic phenomena with spins $S>1/2$, where optical traps allow for the cooling and manipulation of all of the internal hyperfine states of the atom, thus realizing atomic gases with a large spin (e.g. $^{52}$Cr with $S=3$)~\cite{Kawaguchi-2012,2013_SKurn_Ueda_RMP}. 
This can lead to superfluids with non-trivial magnetic structure that spontaneously break both charge conservation and spin rotation symmetries~\cite{Ho-1998,Ohmi-1998}. 

\begin{figure}
\includegraphics[scale=.2]{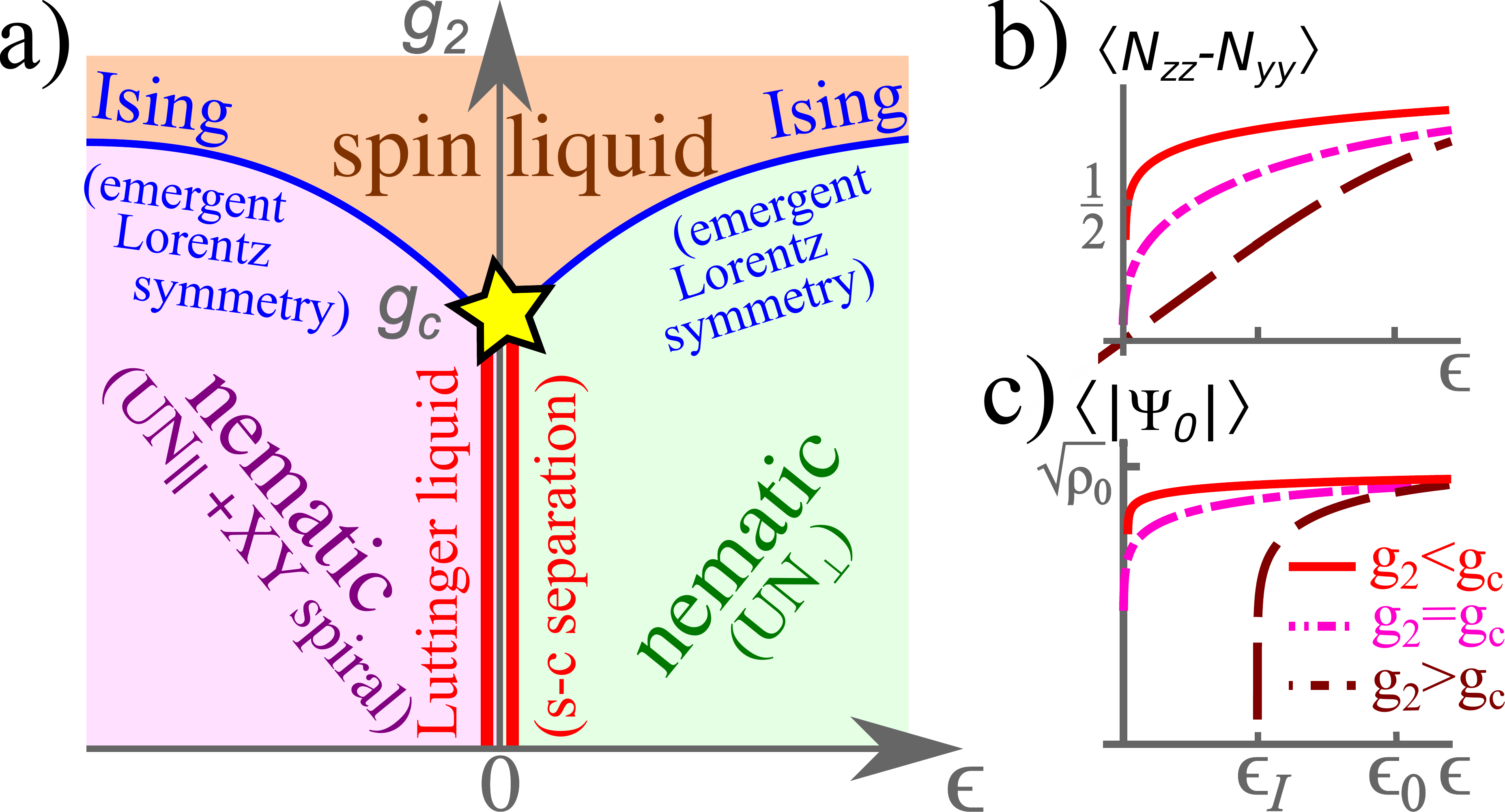}
\caption{a) Phase diagram in the plane spanned by effective quadratic Zeeman field $\epsilon = q + \Theta^2/(2m)$ and spin-spin interaction $g_2$. 
For explanations on the two nematic phases 
and the spin liquid see the main text. The non-universal position $g_c$ of the BKT transition is marked by a star. 
b) Difference of the only non-zero nematicity tensor components $\langle N_{zz} -  N_{yy} \rangle$, note that it is odd in $\epsilon$ and $\langle N_{yy}+ N_{zz} \rangle = 1$. 
The characteristic power law 
is non-universal $\vert \epsilon \vert^{1/(2K_s-1)}$, $K_s\ge 2$ for $g_2 \le g_c$, 
and linear for $g_2 > g_c$. c) The $m_z = 0$ component of the BEC wave function (the order paramater) scales as $ \epsilon ^{(1/4)/(2K_s-1)}$ for $g_2 \leq g_c$ 
and $( \epsilon - \epsilon_I )^{1/8}$ for $g_2 > g_c$. 
}
\label{fig:Phasediagram}
\end{figure}

Ultra cold spin-1 bosons are an ideal system to study nontrivial magnetism beyond conventional vector magnetic order parameters. A pivotal microscopic ingredient is the spin dependent interaction $g_2$ 
which can either be ferromagnetic ($g_2<0$) or polar $(g_2>0)$~\cite{2013_SKurn_Ueda_RMP} and leads to different ground states displaying either non-zero or zero spin expectation value, respectively~\cite{Ho-1998,Ohmi-1998}. In the following, we concentrate on the polar case which is readily realized with $^{23}$Na gases~\cite{2013_SKurn_Ueda_RMP}. The condensate wavefunction can be written as a three-component spinor $\Psi_{\mathrm{MF}} = \sqrt{\rho}e^{i\vartheta}\hat{n}$ where the superfluid phase $\vartheta$ and the unit vector $\hat n$ parametrize the ground state manifold. 
The polar condensate has nematic order signaled by non-zero eigenvalues of a rank-2 tensor order parameter~\cite{Ho-1998,Ohmi-1998}. A quadratic Zeeman field~\cite{Gerbier-2006} 
lifts the degeneracy and the ground state spinor is given by either $\hat{n}=(0,1,0)^T$ or a planar state $\hat{n}=(e^{i \varphi},0,e^{-i \varphi})^T$ depending on the sign of the quadratic Zeeman field~\cite{Kawaguchi-2012}. 
  In recent experiments, it has been demonstrated that it is possible to observe the non-trivial nematic order in $^{23}$Na~\cite{Zibold-2016} and
that the quadratic Zeeman effect can be used to drive nematic phase transitions~\cite{Jacob-2012,Frapolli-2017}. 
Moreover, the nematic planar phase is interesting due to the different types of topological defects that can result from the winding of the phase $\vartheta \rightarrow \vartheta +2\pi$ or the combined operation of a \emph{half}-winding of the phase $\vartheta \rightarrow \vartheta + \pi$ and an inversion of the spinor $\hat{n} \rightarrow -\hat{n}$ that leave $\Psi_{\mathrm{MF}}$ unchanged~\cite{Mukerjee-2006,Podolsky-2009,JamesLamacraft2011}, which have recently been observed in $^{23}$Na~\cite{Seo-2015}. 

With the latest development of artificial gauge fields, it is now possible to couple the internal spin states of the atom to their momentum using counter-propogating Raman lasers, which induces an effective spin orbit coupling (SOC) 
~\cite{Galitski-2015}. 
SOC'ed quantum gases can now be realized in spinor bosons~\cite{Lin-2009,Lin-2011,Stuhl-2015,Campbell-2016,Valdes-2017} or spinful fermions~\cite{Wang-2012} with either a one or two dimensional SOC~\cite{Huang-2016, Wu-2016, Song-2017, Sun-2017}. 
In 
bosonic gases this gives rise to 
``striped'' superfluids~\cite{Ueda-12,Li2012,Li2013,Hickey2014,Martone-2014,Lan2014,Cole2015,Pixley-2016,Hurst-2016,Martone-2016} that condense at the degenerate momenta dictated by the spin orbit wave vector. 
 While the phase diagram is now reasonably well understood,
SOC'ed, polar, spin-1 gases offer an exciting platform to study the competition between different types of nematic order, 
and  hold great promise for intriguing forms of quantum criticality.
 A majority of the theoretical~\cite{Li2012,Lan2014,Cole2015,Pixley-2016,Hurst-2016,Pixley-2017,Cole-2017} and experimental~\cite{Lin-2009,Lin-2011,Stuhl-2015,Campbell-2016,Valdes-2017} work 
 has focused on quantum phase transitions (QPTs) that are driven by the strength of the Raman field and are accessible in both pseudospin-1/2 and spin-1 bosons. Interestingly, for polar spin-1 bosons, the 
 phenomena and nematic QPTs that can be evoked by SOC goes beyond transverse field induced transitions, and remains largely unexplored apart from mean field (MF)~\cite{Hurst-2016,Yu-2016} and variational solutions~\cite{Cole2015,Martone-2016,Sun-2016}. Our work aims to fill this gap by developing a field theory description of nematic QPTs.

One major difficulty in theoretically capturing the interplay between non-perturbative topological defects, SOC, and nematic order is that it requires a strong coupling solution beyond any MF like description. 
Thus one of the most felicitous realms to study SOC'ed polar spinor bosons are one-dimensional (1D) systems, which represents a common setup for ultra cold atom experiments. This is due to the existence of
strong analytical tools that allow for asymptotically exact low-energy solutions that take into account both the inherent strong coupling nature of 1D and topological defects~\cite{Zhou-2001,ZhaiZhou2005,Essler-2009}.
The effective field theory of polar spin-1 bosons in the absence of a SOC is described by a spin-charge separated Lagrangian, the charge is described by a gapless Luttinger liquid (LL) and the spin sector is given by a 1D non-linear sigma model (NL$\sigma$M)~\cite{Zhou-2001,Essler-2009}. A SOC directly couples the spin and charge degrees of freedom and therefore it is in no way obvious if spin-charge separation can still persist in SOC'ed gases.

\textit{Summary of results and experimental predictions.} We consider a gas of 1D polar spinor bosons in the presence of a SOC (wave vector $\Theta$) and a linear (quadratic) Zeeman field $hp$ ($q$). We treat the strength of background fields perturbatively and derive the effective low energy field theory that describes a LL coupled to a NL$\sigma$M in the presence of anisotropic mass terms. We solve this effective theory in the low energy limit and determine the phase diagram of the model, see Fig.~\ref{fig:Phasediagram}. We uncover three distinct superfluid phases: at weak coupling, two different nematic phases 
depending on the sign of the effective quadratic Zeeman field $\epsilon = q + \Theta^2/(2m)$
and 
 a spin liquid phase 
 at strong coupling. Furthermore, we determine the nature of the QPTs between those phases, all of which are continuous. The critical state between the two nematic phases at weak coupling is a pair of spin-charge separated Luttinger liquids. 
In contrast, the transition from either nematic phase to the spin liquid is in the 1+1D Ising universality class with an exotic, emergent Lorentz symmetry characterized by equal, logarithmically divergent velocities in the spin and charge sector. Interestingly, a very similar QPT was discussed  in the physically unrelated context of Cooper pairing near Lifshitz transitions and in topological superconductors~\cite{SitteGarst2009,AlbertonAltmann2017,KaneHalperin2017}. Finally, Ising and LL QPT lines meet at a Berezinskii-Kosterlitz-Thouless (BKT) critical point.

The hallmarks of our theory are as follows: (i) The described phases and fluctuation induced continuous QPTs. We emphasize, that mean field (MF) and variational theories predict a first order transition at $\epsilon = 0$ and miss the spin liquid phase completely. (ii) The order parameter of the QPTs are the spin components of the condensate wave function, see Fig.~\ref{fig:Phasediagram} c). 
(iii) An experimentally accessible observable is the nematic tensor $N_{ab} = \delta_{ab} - \lbrace S_a, S_b \rbrace/2$, see Fig.~\ref{fig:Phasediagram} b). We predict a characteristic power law behavior of $N_{yy},N_{zz}$ with non-universal exponents. This emblematic feature of LL physics is out of reach of MF theory. 
For parameters in typical ultra-cold atom experiments with quasi-1D tubes of atoms at nano-Kelvin temperatures we estimate $K_s\sim \mathcal O(10)$,  and a system size and thermal length which exceed the correlation length~\cite{SuppMat}. Thus, these power-laws should be experimentally detectable.
(iv) The effect of SOC is twofold: First, the condensate wave function in the nematic $\epsilon <0$ phase is heavily modulating in space. Second, SOC strongly affects the position of QPTs. However, somewhat strikingly, the universal critical behaviors are independent of the SOC. (v) Finally, the emergent Lorentz symmetry at the Ising transitions is, at least in principle, accessible via separate measurement of excitation spectra in charge and spin sectors~\cite{MartiStamperKurn2014,BaillieBlakie2016}.
In the remainder we present the theoretical framework leading to these results and predictions.

\textit{Model}: Continuum spin-1 bosons with mass $m$ that are perturbed by a background helical magnetization {and a constant linear Zeeman field} $\vec h(x) = h (\cos (\Theta x), -\sin(\Theta x), p)^T$ as well as a quadratic Zeeman coupling $q$ can be described by the normal ordered Hamiltonian density $\mathcal{H} =  \partial_x \Psi^\dagger \partial_x \Psi/(2m) + \mathcal{H}_2 +  \mathcal{H}_4 $, where
\begin{subequations}
\begin{eqnarray}
\mathcal{H}_2 &=&  {q \Psi^\dagger S_z^2 \Psi }+ { \Psi^\dagger \vec h(x) \cdot \vec S \Psi}, \label{eq:H2} \\
\mathcal{H}_4 &=& \frac{g_0}{2} :(\Psi^\dagger \Psi)^2:+\frac{g_2}{2} :(\Psi^\dagger \vec S \Psi)^2:. \label{eq:H4}
\end{eqnarray}
We analyze the polar case $g_0>g_2 >0$ ($g_0 \sim 32 g_2$ in $^{23}$Na~\cite{2013_SKurn_Ueda_RMP}) in the semiclassical limit in which the condensate density $\rho_0 = \mu/g_0$ parametrically exceeds the inverse coherence length $1/\xi_c = \sqrt{2m \mu}$. Here, $\mu$ is the chemical potential and we set $\hbar=k_B=1$ throughout. 

The bosonic field operators $\Psi, \Psi^\dagger$ are three-spinors and in the remainder we choose the adjoint representation of $\mathbf{SU}(2)$ as a basis of spin-1 operators $(S_a)_{bc} = - i \epsilon_{abc}$ with $a,b,c \in \lbrace x,y,z \rbrace$. 
The quartic term can be recast into the form
$
\mathcal{H}_4 = (g_0 + g_2)/2 :(\Psi^\dagger \Psi)^2:-g_2/2 :\left [\Psi^\dagger   \Psi^* \right ]\left [ \Psi^T  \Psi \right ]:
$
\label{eq:H0}
\end{subequations}
so that the $[\mathbf U(1)\times \mathbf O(3)]/\mathbb Z_2$ symmetry of the unperturbed action becomes manifest. 
Eq.~\eqref{eq:H0} describes the quantum fluid in the lab frame, 
the
frame co-rotating with the Raman field, can be 
accessed by 
$\Psi \rightarrow e^{i \Theta x S_z} \Psi$. 
In this frame, Eq.~\eqref{eq:H0} retains its structure, except for $\vec h \rightarrow  h(1,0,p)^T$ and $\partial_x \rightarrow \partial_x + i \Theta S_z$ (this  yields $q\rightarrow \epsilon= q+\Theta^2/2m$).

\begin{figure}
\includegraphics[scale=.65]{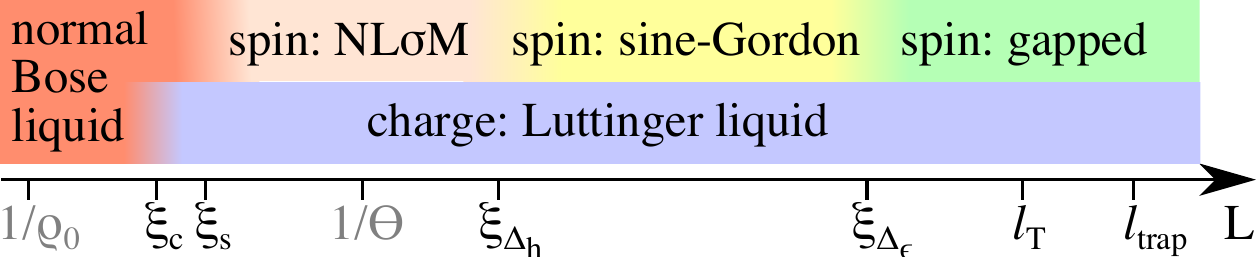} 
\caption{Length scales of the problem away from criticality. The large superfluid density and the slow SOC pitch $\rho_0 \gg 1/\xi_{c,s} \gg \Theta$ enable the controlled derivation of Eq.~\eqref{eq:NLSMLL}. The perturbative inclusion of the effective fields $\vert \epsilon \vert, h \ll \mu$, implies $\xi_{c,s} \ll \xi_{\Delta_h} < \xi_{\Delta_\epsilon} $ (the last inequality reflects the focus on SOC). At each length scale $\xi_{c,s,\Delta_h,\Delta_\epsilon}$, certain modes freeze and an effective theory emerges.}
\label{fig:Lengthscales}
\end{figure}

In order to solve Eq.~\eqref{eq:H0} in its low-energy sector, we perform a sequence of coarse graining steps which are motivated by the assumption of the hierarchy of length scales presented in Fig.~\ref{fig:Lengthscales}. The meaning of each of those scales will be explained at the appropriate position of the main text. Since the dispersion relation of collective modes is linear, see Eq.~\eqref{eq:NLSMLL} below, the conversion to equivalent time (energy) scales follows trivially. 

\textit{Effective low-energy theory.} As a first step towards the asymptotic solution of Eq.~\eqref{eq:H0} we derive the effective long-wavelength Matsubara field theory~\cite{Zhou-2001,Essler-2009}, for details see Ref.~\cite{SuppMat}.
It is convenient to choose an Euler angle parametrization
$
\Psi = \sqrt{\rho} e^{i \vartheta} {O } e^{i \alpha_4 \lambda_4} e^{i \alpha_6 \lambda_6} \hat e_z,
$
with $\lambda_i$ being Gell-Mann matrices. This representation separates the Goldstone modes $e^{i \vartheta}, O = e^{i \alpha_7 \lambda_7} e^{i \alpha_5 \lambda_5}$ living on the manifold $[\mathbf U(1) \times \mathbf O(3)/\mathbf O(2)]/\mathbb Z_2$ from the massive longitudinal modes $\alpha_4$ and $\alpha_6$ from the outset. 
This representation of the complex unit vector $\Psi/\sqrt{\rho}$ provides a regular Jacobian leading to the NL$\sigma$M measure for the Goldstone field $\hat n \equiv O \hat e_z \in \mathbb S^2$. While constant $\vartheta$ and $O$ fields are zero modes of $\mathcal H - \mathcal H_2$,
Eqs.~\eqref{eq:H2},\eqref{eq:H4} ensure that the longitudinal modes take the saddle point values
$\rho_{\rm MF}=\rho_0-q \hat n S_z^2 \hat n/g_0$, $\alpha_{\rm 4,MF}=-i \hat e_z O^T \vec h \cdot \vec S O \hat e_x/[2\rho_0 g_2]$ and $\alpha_{\rm 6,MF}=-i \hat e_z O^T \vec h \cdot \vec S O \hat e_y/[2\rho_0 g_2]$, which are perturbative in $h g_0/(\mu g_2)$ but non-perturbative in $q$.
Fluctuations around the saddle point 
$\Delta \rho$ ($\Delta \alpha_{4,6}$) 
decay on the length scale $\xi_c$ ($\xi_s = \sqrt{g_0/g_2} \xi_c$). To access the physics at longer scales, we perform the Gaussian integration of massive modes assuming that $O$ and $\vartheta$ are slow. 
We switch to the co-rotating frame and obtain the effective low-energy Lagrangian $\mathcal L =  \mathcal L_0 + \mathcal L_1 + \mathcal L_2$, 
\begin{subequations}
\begin{align}
\mathcal L_0 &= \Delta_\epsilon \hat n S_z^2 \hat n - \Delta_h\hat n (S_x+p S_z)^2 \hat n , \label{eq:PotTermsAction}\\
\mathcal L_1 &=   - i \dot \vartheta \lambda_\epsilon \hat n S_z^2 \hat n + \lambda_h \dot{\hat n}   S_x \hat n + i \lambda_\Theta \hat n' S_z \hat n, \\
\mathcal L_2 &=  \frac{K_c}{2\pi v_c} \left [{\dot \vartheta}^2 +{v_c^2}{\vartheta'}^2 \right] +  \frac{K_s}{2\pi v_s}\left [ \vert\dot{ \hat n} \vert^2 + v_s^2 \vert \hat n'\vert^2 \right  ] . \label{eq:KinTermsAction}
\end{align}
\label{eq:NLSMLL}
\end{subequations}
The kinetic part of the action, Eq.~\eqref{eq:KinTermsAction}, which we denote as $\mathcal{L}_2 = \mathcal{L}_{\rm LL}[\vartheta] + \mathcal{L}_{\rm NL\sigma M}[\hat n]$,
 contains bare coupling constants $K_{c,s} = \sqrt{2} \pi {\rho_0}\xi_{c,s}$ and velocities $v_{c} = \sqrt{{\rho_0 g_{0}}/{m}}$ and $v_{s} = \sqrt{{\rho_0 g_2}/{m}}$. We omitted anisotropic corrections to kinetic terms due to $q, \Theta$ and $h$, because they are small and will
renormalize to zero quickly. In addition to the known kinetic term $\mathcal L_2$, Eq.~\eqref{eq:NLSMLL}
contains
symmetry breaking terms with no derivatives $\Delta_\epsilon =  \rho_0 \epsilon, \Delta_h = {h^2}/{2g_2}$ and one derivative $\lambda_\epsilon = {\epsilon}/{g_0} , \lambda_h ={h}/{g_2}, \lambda_\Theta ={\Theta \rho_0}/{m}$ which are the focus of this letter. 
{In Ref.~\cite{SuppMat} we treat a weak trapping frequency $\omega_\Vert \ll m g_0^2$ via the replacement $\rho_0 \rightarrow \rho_0 [1 -x^2/l_{\mathrm{trap}} ^2]$. We find that this introduces the largest finite length scale ($l_{\mathrm{trap} } = \sqrt{2\mu/m\omega_\Vert ^2})$ into the problem, which is less restrictive then the presence of finite temperature $(l_T = v_s/T$), and their combined effect rounds out the observable critical properties (see Fig.~\ref{fig:Lengthscales}).}

\textit{Characterization of phases.} We begin the asymptotic solution of Eq.~\eqref{eq:NLSMLL} by determining all phases and their characteristics, see Fig.~\ref{fig:Phasediagram} a). Groundstates which are also accessible to variational~\cite{Cole2015,Martone-2016,Sun-2016} and MF~\cite{Hurst-2016,Yu-2016} treatments follow from the consideration of the potential term $\Delta_\epsilon  S_z^2 - \Delta_h (S_x+pS_z)^2$ which independently of $p$ predicts a first order transition at $\epsilon = 0$~\cite{SuppMat}. For $p = 0$ it has eigenvalues $\Delta_\epsilon, \Delta
_\epsilon - \Delta_h, - \Delta_h$ with eigenstates $\hat e_x, \hat e_y, \hat e_z$, respectively {(for $p\neq 0$ see~\cite{SuppMat})}. 
At finite $h$, the groundstate at $\epsilon >0$ ($\epsilon <0$) is $\Psi_{\mathrm{MF}} \simeq\sqrt{\rho_0} e^{i \vartheta}[\hat e_z + h \hat e_y/(2g_2 \rho_0)]$ ($\Psi_{\mathrm{MF}} \simeq \sqrt{\rho_{\rm MF}} e^{i \vartheta}[\hat e_y - h \hat e_z/(2g_2 \rho_0)]$), where the finite $h$ corrections stem from $\alpha_{4,6}^{\rm MF}. $
This state is denoted UN$_\perp$ (UN$_\Vert$ + XY spiral) because at MF level it displays uniaxial nematic order $\langle N_{zz} \rangle = \rho_0 + \mathcal O(h^2)$ ($\langle N_{yy} \rangle = \rho_0 + \mathcal O(h^2)$).
Both states show weak magnetization 
$\langle S_x \rangle = - h/g_2$. In the lab frame the magnetization follows the helical magnetic field and for $\epsilon <0$ there is a strong modulation of the superfluid wavefunction because bosons condense at finite momentum $k=\Theta$ producing a stripe superfluid~\cite{SuppMat}.
MF theory predicts a first order transition at $\epsilon = 0$: the ground state in the spin sector becomes degenerate and the order parameter $\langle N_{ab}\rangle $ changes discontinuously. 
Finally, there is a third phase in which the spin sector is quantum disordered, i.e. a spin liquid \cite{Essler-2009}. This occurs when $K_s \rightarrow 0$, a scenario that is not captured by the bare parameters entering Eq.~\eqref{eq:NLSMLL} but can be reached upon RG transformations.

\textit{Characterization of phase transitions.} Having identified the three phases of the problem, we now characterize the nature of the QPTs between them. 
We first discuss the RG flow close to the repulsive fixed point $K_s = \infty$ at small $\Delta_{\epsilon, h}$, $\lambda_{\epsilon, h, \Theta}$. It is well known that 
$
d K_s/d b = -1/2 + \mathcal O\left (1/K_s,\Delta_{\epsilon, h}, \lambda_{\epsilon, h, \Theta}\right ).
$
As usual, $b$ denotes the running logarithmic scale.
The unperturbed weak coupling theory suggests that the spin liquid is approached at the length scale $\xi_{SL} \sim \xi_s \exp(\sqrt{2} \pi \rho_0 \xi_s)$. However, the scaling dimensions of $\Delta_{\epsilon,h}$, $\lambda_\epsilon$, and $\lambda_{h, \Theta}$ are $[2-3/(2K_s)]$, $[1-3/(2K_s)]$, and $[1-1/(2K_s)]$, i.e. RG relevant at weak coupling. We define the length scales $\xi_{\Delta_{\epsilon,h}}$ self consistently as the scale when the couplings $\Delta_{\epsilon, h}(b)$ hit the running scale, by assumption $\xi_{\Delta_{h}}<\xi_{\Delta_{\epsilon}}$. Beyond $\xi_{\Delta_h}$ the NL$\sigma$M field is locked to the easy plane $\hat n = (0, \sin (\phi), \cos(\phi))^T$ perpendicular to the background magnetization realizing a spin-flop-like phase of itinerant polar bosons. Following Fig.~\ref{fig:Lengthscales} a sine-Gordon theory emerges. The coupling to the charge Luttinger liquid is characterized by $\mathcal{L}_{\rm EP}=\mathcal{L}_{\rm LL}[\vartheta]+\tilde{\mathcal{L}}$,
\begin{equation}
\tilde{\mathcal{L}} =   \frac{K_s}{2\pi v_s}\left [{(\dot \phi)}^2 + v_s^2 (\phi')^2 \right ] +\left [\Delta_\epsilon   - i \dot \vartheta \lambda_{\epsilon} \right] \sin^2(\phi) .\label{eq:SineGordon}
\end{equation}
All coupling constants in Eq.~\eqref{eq:SineGordon} are evaluated at the scale $\xi_{\Delta_h}$ and we absorbed a factor of $1/(1+p^2)$ into $\Delta_\epsilon, \lambda_\epsilon$. Note that, while $K_c \gg 1$ by assumption, $K_s$ is large only if $\xi_{\Delta_h} \ll \xi_{\rm SL}$ and may be renormalized to values of the order of unity or even smaller otherwise.
In terms of Eq.~\eqref{eq:SineGordon}, the phase UN$_\perp$ (UN$_\Vert$ + XY spiral) is characterized by $\langle \phi \rangle = 0 \text{ mod } \pi $ ($\langle \phi \rangle = \pi/2 \text{ mod } \pi$). 

\begin{figure}
\includegraphics[scale=.32]{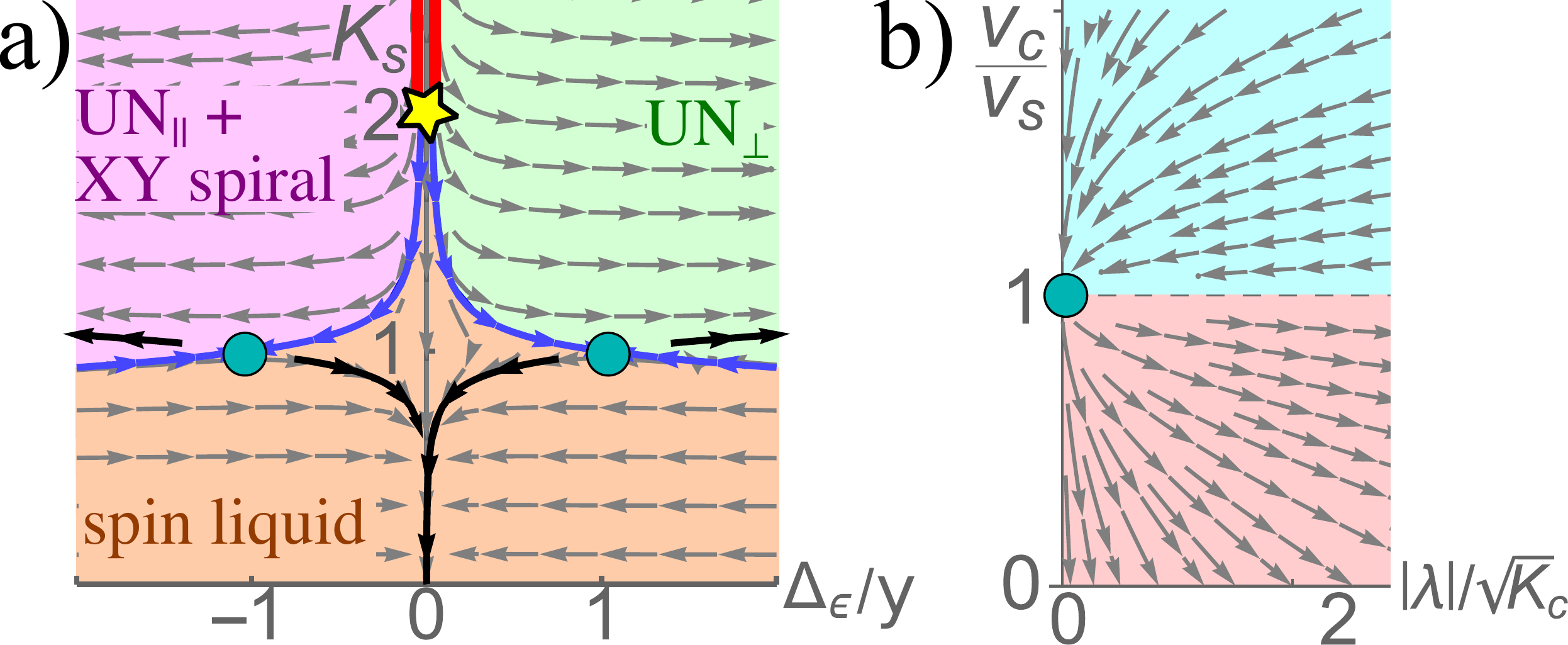} 
\caption{Panel a): RG flow according to Eq.~\eqref{eq:RG_JoseNelson_Extended} in the plane $\Delta_\epsilon y = 0.01$ (color coding as in Fig.~\ref{fig:Phasediagram}). 
The BKT critical end point (Ising fixed point) is represented as a yellow star (turquoise disc). The Ising point resides at $\Delta_\epsilon y = \infty$, and controlled RG equations unveiling its emergent Lorentz symmetry, Eq.~\eqref{eq:IsingBoseRG}, are plotted in panel b).}
\label{fig:RGFlow}
\end{figure}

The fields entering Eq.~\eqref{eq:SineGordon} allow for various topological defects: $2\pi$ phase slips in $\vartheta$ and $\phi$ fields as well as $\pi$ phase slips in $\vartheta$ accompanied with a $\pm \pi$ phase slip in $\phi$ \cite{Mukerjee-2006}. The scaling dimensions~\cite{KrugerScheidl-2002,Podolsky-2009,SuppMat} of the associated fugacities (Boltzmann weights) are $(2- K_c)$, $(2 - K_s)$ and $[2 - (K_c + K_s)/4]$, respectively. Therefore, in the given parameter regime $(K_c \gg 1)$, only the fugacity $y$ of $2\pi$ phase slips in the spin field $\phi$ may be relevant. We incorporate the associated operator into Eq.~\eqref{eq:SineGordon} and derive \cite{SuppMat} the weak coupling RG equations to second order in $\lambda_\epsilon, \Delta_\epsilon, y$ and to zeroth order in $1/K_c$ extending the previously reported~\cite{JoseNelson1977} results to the case of finite $\lambda_\epsilon$: 
\begin{align}
\frac{d \Delta_{\epsilon}}{d b} &= (2 - 1/K_s) \Delta_{\epsilon}, & \frac{d  y }{d b}&= (2 - K_s) y, \notag \\
\frac{ d K_s}{d b} &= \Delta_{\epsilon}^2 - K_s^2 y^2, & \frac{  d\lambda_\epsilon }{d b}& = (1 - 1/K_s) \lambda_\epsilon, \notag \\
\frac{ d (K_c/v_c)}{d b} &= \frac{\lambda_\epsilon^2}{K_s v_s}, & \frac{ d (K_c v_c)}{d b} &= \frac{ d v_s}{d b} = 0 . \label{eq:RG_JoseNelson_Extended}
\end{align}
Regularization dependent factors were absorbed into a redefinition of $\lambda_\epsilon, \Delta_\epsilon, y$. Figure~\ref{fig:RGFlow} a) displays the RG flow in the plane $(\Delta_{\epsilon} / y , K_s)$ and illustrates that (i) the MF first order transition at $\epsilon = 0$ for $K_s \geq 2$ is actually continuous and described by a line of spin-charge seperated LL critical points with enhanced symmetry, (ii) the phase transition to the spin disordered phase is BKT at $\epsilon = 0$, and (iii) the quantum critical point at $\epsilon \neq 0$ occurs at $K_s = 1$, but at strong coupling $\Delta_\epsilon, y \rightarrow \infty$. At this fixed point, the spin charge coupling $\lambda_\epsilon$, which is relevant (irrelevant) for $K_s >1$ ($K_s < 1$), becomes marginal. 
To determine the relevance of $\lambda_\epsilon$ and the nature of the strong coupling phase transition, Eq.~\eqref{eq:SineGordon} is fermionized~\cite{Ogilvie1981,SuppMat} on the $K_s = 1$ hyperplane leading to $\mathcal{L}_{\rm EP,K_s=1}=\mathcal{L}_{\rm LL}[\vartheta]+\mathcal{L}_{\rm F}$
\begin{eqnarray}
\mathcal {L}_{\rm F} &=&  \frac{1}{2}\eta^T \left [\partial_\tau + v_s \hat p \sigma _z + (M_\epsilon + i \lambda \dot \vartheta)\sigma_y \kappa_z + M_v \sigma_y\right ] \eta. \,\,\,\,\,\,\,\, 
\label{eq:IsingBose}
\end{eqnarray} 
The Majorana four spinor $\eta$ is subject to masses $M_\epsilon \sim \Delta_\epsilon \xi_s, M_v \sim y \xi_s$ and coupled to the bosonic charge field via $\lambda \sim \lambda_\epsilon \xi_s$. Pauli matrices in left-right (Nambu) space are denoted $\sigma_a$ ($\kappa_a$). At $\lambda = 0$, two Ising transitions occur at $M_\epsilon = \pm M_v$, corresponding to the turquoise discs in Fig.~\ref{fig:RGFlow} a). The effective theory, Eq.~\eqref{eq:IsingBose}, at the critical point corresponds to a single gapless Majorana mode coupled to a gapless boson by a Lorentz symmetry breaking term. This effective theory is related to the problem studied in Refs.~\cite{SitteGarst2009,AlbertonAltmann2017,KaneHalperin2017} by means of a Lorentz boost $(v_s \tau, x) \rightarrow (x, -v_s \tau)$ and an analytical continuation $\lambda \rightarrow i \lambda$. In that case, an attractive weak coupling fixed point $\lambda \rightarrow 0$ with emergent Lorentz symmetry and vanishing velocity $v_c = v_s \rightarrow 0$ was uncovered along with a putative phase separated region at strong coupling. Returning to our theory, it is useful to present the one-loop RG equations in terms of 
$G = {\vert \lambda \vert}/{\sqrt{K_c}}, u = {v_c}/{v_s}, \bar v =  \sqrt{v_c v_s}$
\begin{align}
\frac{d G}{d b} &= \frac{u G^3}{8} \frac{(1-u)(3+u)}{(1+u)^2}, &
\frac{d u}{d b} &= - \frac{u^2 G^2}{4}\frac{(1- u)^2}{(1+u)^2}, \notag \\
\frac{d \bar v}{d b} &= \frac{u \bar v G^2 }{8  }\frac{10u - u^2-1}{(1+u)^2}, &
\frac{d K_c}{d b} &= \frac{u G^2}{4 }K_c. \label{eq:IsingBoseRG}
\end{align}
The mass has scaling dimension $1+ {u G^2( u +1/2)}/{(1+u)^2}$.
Due to the imaginary coupling in our model, the flow is reversed as compared to Refs.~\cite{SitteGarst2009,AlbertonAltmann2017,KaneHalperin2017}, hence $\bar v$ increases near $u = 1$. The first two RG equations in Eq.~\eqref{eq:IsingBoseRG} decouple and are plotted in Fig.~\ref{fig:RGFlow}, b). The assumption $g_0 >g_2$ implies starting values $v_c > v_s$, therefore the effective theory~\eqref{eq:IsingBose} resides in the basin of attraction of the weak coupling fixed point $(\lambda, v_c/v_s) = (0, 1)$. By consequence the critical theory separating the spin disordered from the nematic phases at finite $|\epsilon|$ is a theory with central charge $c = 3/2$, emergent Lorentz symmetry $v_c = v_s$, and logarithmically divergent velocity. 

This concludes the derivation of the quantum critical theories. The zero temperature scaling of the order parameter and nematic tensor, Fig.~\ref{fig:Phasediagram}, is weakly rounded at 
{finite temperature in the center of a harmonic trapping potential} and obtained via a semiclassical evaluation using renormalized coupling constants~\cite{SuppMat}. 
In particular, the semiclassically expected first order jump is washed out by the strong quantum fluctuations at $\epsilon = 0$ which corroborates the significance of the quantum field theoretical analysis. 
{It will be interesting to study the predicted QPT numerically using the density matrix renormalization group to solve the SOC spin-1 Bose-Hubbard model~\cite{Pixley-2016}. Despite the SOC removing any spin conserving quantum numbers~\cite{Pixley-2017}, we expect a numerical solution remains tractable in the superfluid regime provided that the truncation of the bosonic Hilbert space is treated carefully~\cite{Cole-2017}.}

\acknowledgements{\textit{Acknowledgements} We acknowledge useful discussions with I. Bloch,  B. J. DeSalvo, Y. Komijani, J. Lee, P. P. Orth, A. Rosch, A. M. Tsvelik, and J. Wilson. Work by E.J.K. was supported by the U.S. Department of Energy, Office of Science, Office of Basic Energy Sciences, under Award DE-FG02-99ER45790. J.H.P. acknowledges the Aspen Center for Physics where some of this work was performed, which is supported by National Science Foundation grant PHY-1607611.}

%

\newpage
\begin{widetext}

\setcounter{equation}{0}
\setcounter{figure}{0}
\setcounter{table}{0}
\setcounter{page}{1}
\makeatletter
\renewcommand{\theequation}{S\arabic{equation}}
\renewcommand{\thefigure}{S\arabic{figure}}
\renewcommand{\bibnumfmt}[1]{[S#1]}
\renewcommand{\citenumfont}[1]{S#1}

\begin{center}
Supplementary materials on \\
\textbf{"QUANTUM FIELD THEORY OF NEMATIC TRANSITIONS IN SPIN ORBIT COUPLED SPIN-1 BOSONS"}\\
E. J. K\"onig and J. H. Pixley\\ 
\textit{Department of Physics and Astronomy, Center for Materials Theory, Rutgers University, Piscataway, NJ 08854}
\end{center}

\section{Derivation of Low-energy theory presented in Eq. (2) of the main text.}
As explained in the main text we employ the density-phase parametrization of the wave function $\Psi = \sqrt{\rho} \hat m $.
Here, $\hat m$ is a \textit{complex} unit vector $\hat m^\dagger \hat m =1$. It is instructive to parametrize it using a $\mathbf{U}(3)/\mathbf U(2)$ Euler angle parametrization
\begin{equation}
\hat m = e^{i \vartheta} \underbrace{e^{i \alpha_7 \lambda_7} e^{i \alpha_5 \lambda_5} }_{=: O} \underbrace{e^{i \alpha_4 \lambda_4} e^{i \alpha_6 \lambda_6} \hat e_z}_{= \left (\begin{array}{c}
i \sin(\alpha_4) \cos(\alpha_6) \\ 
i \sin(\alpha_6) \\ 
\cos(\alpha_4)\cos(\alpha_6)
\end{array} \right )}.
\end{equation}

The logic of this parametrization is as follows: generically one can represent $\hat m = U \hat e_z$ with $U \in \mathbf U(3)$ and $\hat e_z = (0, 0, 1)^T$. However, the $\mathbf U(2)$ subgroup leaving $\hat e_z$ invariant has to be divided out. On the level of the tangent space, this is achieved as follows: we split the nine generators in $\mathfrak u(3)$ into two sets $\lbrace{\mathbf 1, \lambda_4, \lambda_6, \lambda_5, \lambda_7}\rbrace$ and $\lbrace{\lambda_1, \lambda_2, \lambda_3, \text{diag}(1,1,0)}\rbrace$. For the quotient space, we simply remove the second set. The fact that this parametrization is covering the whole target space (at least near the mean field saddle point) follows from the calculation of the Jacobian, below.

\textbf{Potential terms.} In the chosen parametrization the potential terms for $h = 0$ and $q = 0$ are
\begin{equation}
- \mu \Psi^\dagger \Psi + \mathcal H^{(4)} = - \mu \rho + \frac{g_0}{2} \rho^2 + \frac{g_2}{2}\rho^2 \left (1 - (1-2 \cos^2(\alpha_4)\cos^2(\alpha_6))^2\right ).
\end{equation}
The mean field solution of this potential is $ \rho = \rho_0 \equiv \frac{\mu}{g_0}, 
(\alpha_4, \alpha_6) \in  \lbrace (0,0),(0, \pi), (\pi, 0), (\pi,\pi) \rbrace.$
It is manifest that $O$ and $\vartheta$ do not enter and are thus zero-modes of the unperturbed theory. For the first and fourth solution of $\alpha_{4,6}$ we have $e^{i \alpha_4 \lambda_4} e^{i \alpha_6 \lambda_6} \hat e_z = \hat e_z$, for the second and third the sign of the vector is reversed. Thus there are only two distinct saddle points, which in addition can be rotated from one to another by means of $O$. Therefore, in what follows, we concentrate on the vicinity of $(\alpha_4, \alpha_6) = (0,0)$, only.

We now restore $h$ and $q$ and follow the semiclassical strategy of finding a saddle point solution perturbatively in $h$ and subsequently perform a Gaussian integration. To this end we first expand the full $\mathcal H^{(2)} + \mathcal H^{(4)}$ up to second order in $\Delta \vec P = (\Delta \rho, \alpha_4,\alpha_6)$ where $\Delta \rho = \rho - \rho_{\rm MF}$ and $\rho_{\rm MF}$ is the constant mean field superfluid density that remains to be determined. This procedure yields
\begin{subequations}
$
\mathcal H^{(2)} + \mathcal H^{(4)} = V_{\rm MF} + \vec A_V \Delta \vec P + \frac{1}{2} \Delta \vec P \underline M^{-1} \Delta \vec P
$ with
\begin{eqnarray}
V_{\rm MF} &=& - \mu \rho_{\rm MF} + \frac{\rho_{\rm MF}^2 g_0}{2} + \rho_{\rm MF} q \hat n S_z^2 \hat n, \label{eq:VMF}\\
\vec A_V &=& \left (\begin{array}{c}
- \mu + \rho_{\rm MF} g_0 + q \hat n S_z^2 \hat n \\ 
2 i \rho_{\rm MF} \hat e_z O^T \vec h\cdot \vec S O\hat e_x \\ 
2 i \rho_{\rm MF} \hat e_z O^T \vec h\cdot \vec S O\hat e_y
\end{array} \right ) , \\
\underline M^{-1} &=& \text{diag}\left (g_0, 4 \rho_{\rm MF}^2 g_2, 4 \rho_{\rm MF}^2 g_2\right ).
\end{eqnarray}
\end{subequations}
We have omitted the $x$-dependence of $\vec h$ for notational convenience. Corrections to $\underline M$ lead to terms that are small in $\lbrace q, h\rbrace/\mu$ and shall be dropped eventually.

In this notation, one can readily determine the saddle point values $\Delta \vec P  = -\underline M \vec A_V$ with $\rho_{\rm MF} = \mu/g_0 - q \hat n S_z^2 \hat n/g_0$ and perform the Gaussian integral by completing the square. 

\textbf{Time derivative term, gradient term and Jacobian.}
Next, we consider the time derivative term that occurs in a Euclidean path integral treatment $
\Psi^\dagger \dot \Psi = i \dot \vartheta \rho_{\rm MF} + \Delta \vec P \vec A_\tau.
$
We omitted terms with time derivatives on massive fields and introduced
\begin{equation}
\vec A_\tau = \left (\begin{array}{c}
i \dot \vartheta \\ 
2i \rho_{\rm MF} \hat e_z O^T \dot O \hat e_x \\ 
2i \rho_{\rm MF} \hat e_z O^T \dot O \hat e_y
\end{array} \right ).
\end{equation}
The gradient term is expanded as
$
\frac{\nabla\Psi^\dagger \nabla \Psi}{2m} = \mathcal H_{\rm kin}\vert_{\rm MF} + \frac{1}{2m} \Delta\vec P' \underline g_{\Delta P}\Delta \vec  P'$, where the prime denotes a spatial derivative on $\Delta\vec P$.
We have introduced
\begin{subequations}
\begin{eqnarray}
\mathcal H_{\rm kin}\vert_{\rm MF} &=& \frac{\rho_{\rm MF}}{2m}\left [\vartheta'^2 + \vert\hat n'\vert^2\right ], \label{eq:HMF} \\
\underline g_{\Delta \vec P}&=& \text{diag}(1/[4\rho_{\rm MF}],\rho_{\rm MF}, \rho_{\rm MF}).
\end{eqnarray}
\end{subequations}
Note that linear in $\Delta \vec P$ terms with two spatial derivatives on $\vartheta$ and  $O$ are omitted because they induce corrections that are parametrically small.
The $\omega = 0$ propagator of massive modes $[\underline g_{\Delta \vec P} \frac{p^2}{2m} + \underline M_0^{-1}]^{-1}$ determines the decay length $\xi_c$ ($\xi_s$) of massive modes $\Delta \rho$ ($\alpha_{4,5}$) (we used $\rho_{\rm MF} = \mu/g_0 + \mathcal O(q)$, here). 
Using the above parametrization of $O$ we obtain $\vert \hat n' \vert^2 = (\alpha_5')^2 + \cos(\alpha_5)^2 (\alpha_7')^2$. The Jacobian can be readily calculated from the metric $\nabla \Psi^\dagger \nabla \Psi = \nabla (\Delta \vec P; \vartheta, \alpha_5, \alpha_7) \underline g \nabla (\Delta \vec P; \vartheta, \alpha_5, \alpha_7)^T$ with $\underline g = \text{diag}\left (1/[4 \rho_{\rm MF}], \rho_{\rm MF}, \rho_{\rm MF}; \rho_{\rm MF}, \rho_{\rm MF}, \rho_{\rm MF}\cos(\alpha_5) \right )$ and is given by 
$
J = \sqrt{\det \underline g} = \cos(\alpha_5) \rho_{\rm MF}^2/2.
$
Shifting $\alpha_5$ by $\pi$ we obtain the standard measure on a sphere. Upon Gaussian integration of $\Delta \vec P$, the factor of $\rho_{\rm MF}^2$ from the Jacobian and fluctuation determinant cancel.

\textbf{Integration of massive modes.} From the integration of massive modes we obtain 
$
\Delta S = - \frac{1}{2}\int_{\tau, x}  (\vec A_V+\vec A_\tau) \underline M (\vec A_V+\vec A_\tau)
$
with the following leading terms in terms of small $h/\mu, q/\mu$
\begin{subequations}
\begin{eqnarray}
- \frac{1}{2} \vec A_\tau \underline M_0 \vec A_\tau &=& \frac{1}{2}\left (\frac{(\dot \vartheta)^2}{g_0} + \frac{(\dot {\hat n})^2}{g_2}\right ) , \\
- \frac{1}{2} \vec A_V \underline M_0 \vec A_V &=& -\frac{1}{2g_2} \hat n (\vec h \cdot \vec S) ^2\hat n ,\\
- \vec A_V \underline M_0 \vec A_\tau &=& - i \dot \vartheta \frac{q}{g_0} \hat n S_z^2 \hat n - i \dot \vartheta \rho_{\rm MF} + \frac{1}{g_2} \dot{\hat n} \vec h \cdot \hat S \hat n.
\end{eqnarray}
\label{eq:Fluctuations}
\end{subequations}
We anticipated that $-\mu + g_0\rho_{\rm MF} = \mathcal O(q)$ and, as mentioned in main text, omitted anisotropic kinetic terms. 

Collecting Eqs.~\eqref{eq:VMF},\eqref{eq:HMF},\eqref{eq:Fluctuations} yields the effective low-energy theory in the lab frame. In order to obtain the equivalent theory in the rotating frame by applying the local rotation $\hat n \rightarrow e^{i \Theta x S_z} \hat n$, which amounts to the replacement $q \rightarrow \epsilon$ everywhere and a single derivative term stemming from Eq.~\eqref{eq:HMF}. This concludes the derivation of Eq. (2) of the main text.

\begin{figure}
  \begin{center}
    \includegraphics[width=0.6\textwidth]{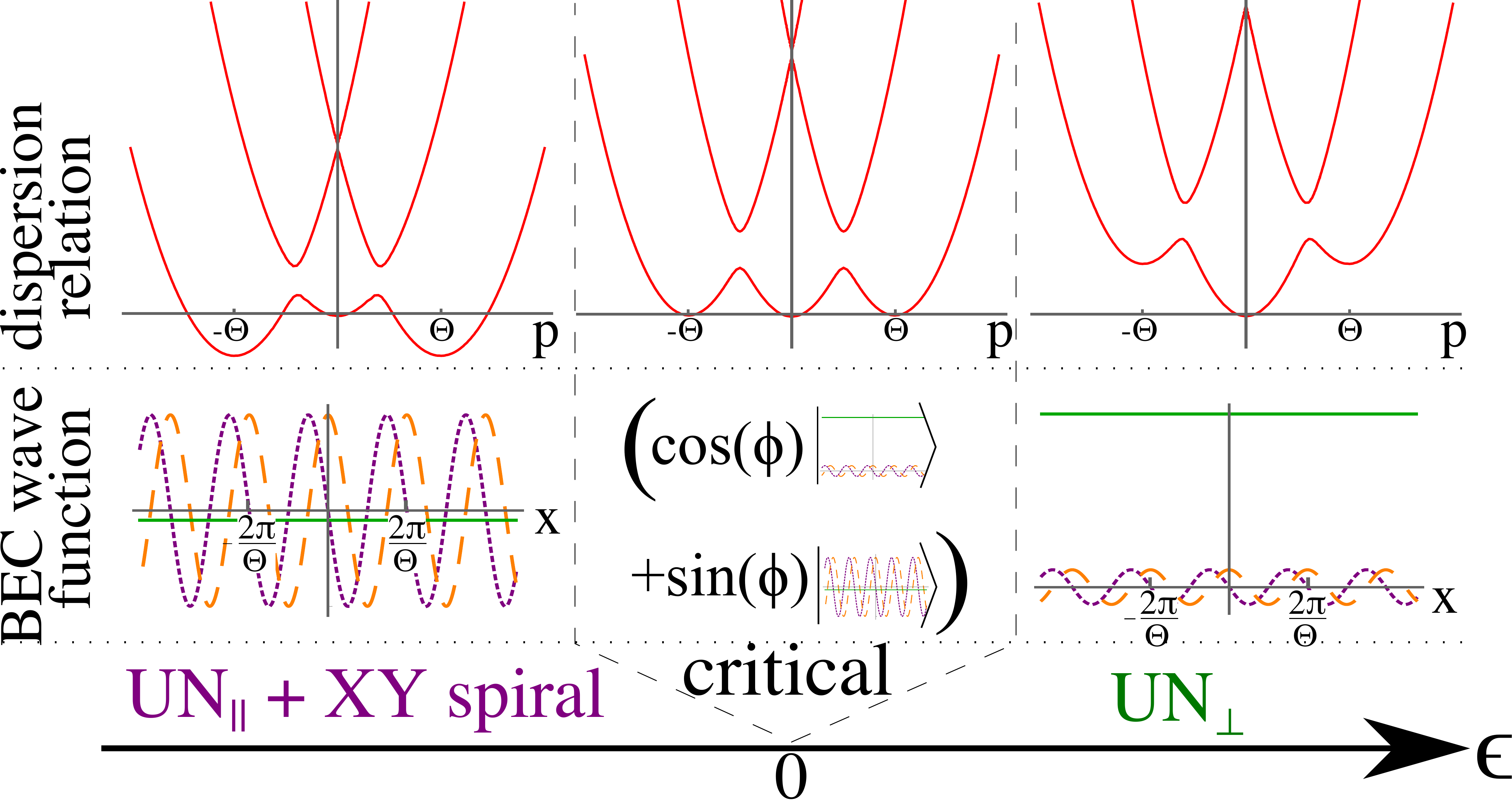} \quad 
    \includegraphics[width=0.35\textwidth]{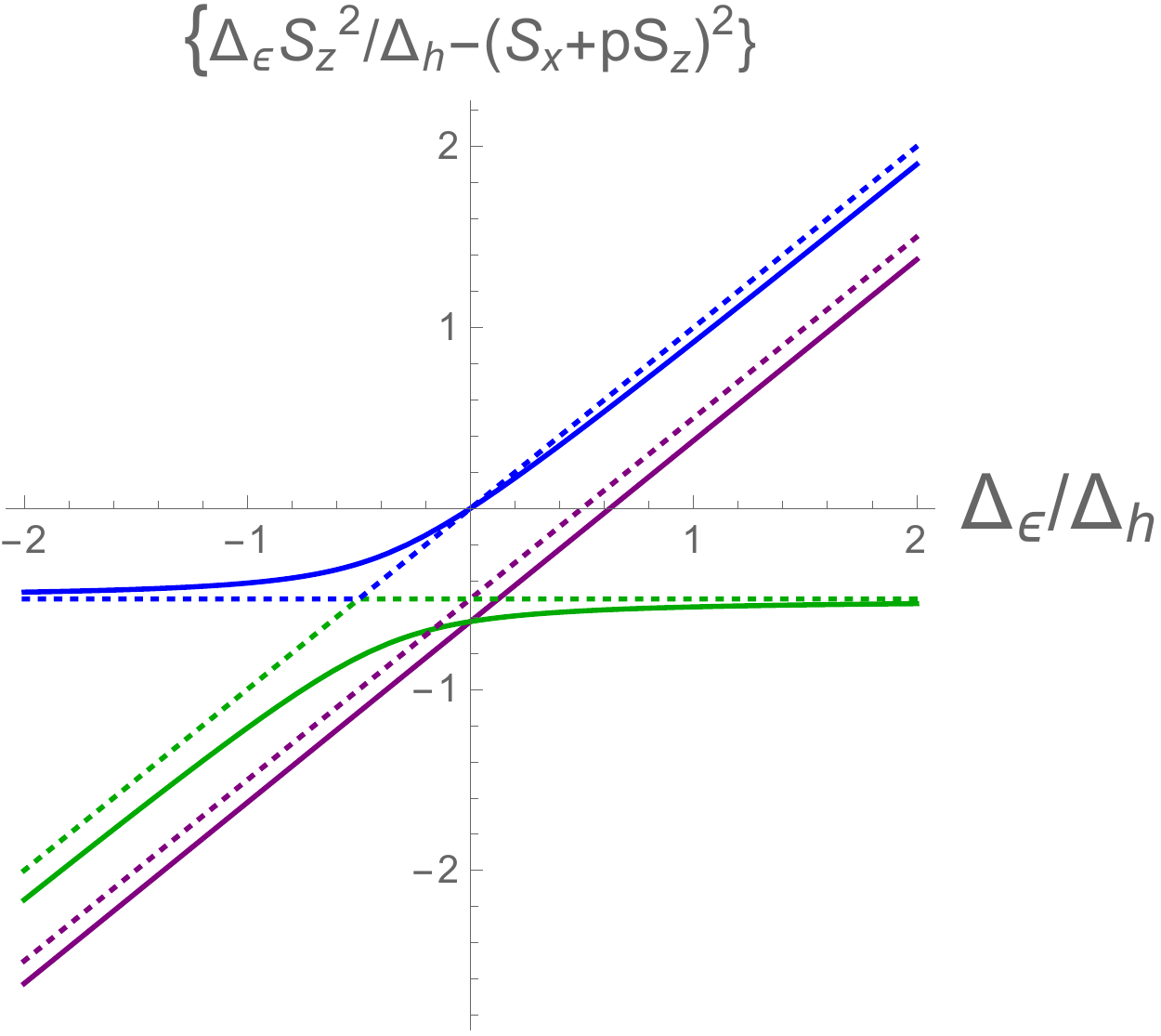} 
  \end{center}
  \caption{Visualization of mean field properties. Left, first row: dispersion relation of non-interacting bosons at $p =0$. Left, second row: mean field wave function. The strong modulation in the spiral phase stems from predominant condensation at $k = \pm \Theta$. The x,y, and z components of the spinor wave functions are represented as dotted purple, dashed orange and solid green curves, respectively, and $h/(2g_2 \rho_0) = 0.1$. Left, third row: Associated phases, as discussed in the main text. Right: Eigenvalues of the matrix $\Delta_\epsilon S_z^2/\Delta_h - (S_x + p S_z)^2$ at $p = 0.5$ (solid) and $p = 0$ (dashed).}
  \label{SM:fig:DispRel}
\end{figure}

\textbf{Mean field phase diagram and wave function.}
To determine the
wave function at the mean field level at first order in $h, q$ (or $\epsilon$) one has to take the saddle point value of $\Delta \vec P$ into account. 
\begin{equation}
\Psi = \sqrt{\rho_{\rm MF} + \Delta \rho} e^{i \vartheta} O e^{i \alpha_4 \lambda_4}e^{i \alpha_6 \lambda_6} \hat e_z \doteq \sqrt{\rho_{\rm MF}} e^{i \vartheta} \hat n - \frac{e^{i \vartheta}}{2 \sqrt{\rho_{\rm MF}}} \frac{\vec h \cdot \vec S}{g_2} \hat n \label{eq:MFWF}
\end{equation}
The symbol $\doteq$ refers to the replacement $\Delta \vec P \rightarrow - \underline M \vec A_V$ under Gaussian $\Delta \vec P$ integration at mean field level. In conclusion the mean field expectation values for an operator $O$ in the rotating wave frame are
\begin{equation}
\Psi^\dagger O \Psi = \rho_{\rm MF} \left [\hat n O \hat n - \frac{h}{2g_2 \rho_0} \hat n \lbrace S_x+p S_z, O \rbrace \hat n\right ], \label{eq:MFObs}
\end{equation}
where have denoted the anticommutation relation by $\{ \dots \}$.
The mean field phase diagram follows from considering the potential $\Delta_\epsilon  S_z^2 - \Delta_h (S_x + pS_z)^2$. It has energies $\Delta_\epsilon - \Delta_h (1+p^2), (\Delta_\epsilon - (1+p^2) \Delta_h \pm \sqrt{(\Delta_h (1+p^2) - \Delta_\epsilon)^2+4 \Delta_\epsilon \Delta_h})/2$ with associated unnormalized eigenvectors $\hat e_y, (\Delta_h (1- p^2) + \Delta_\epsilon \pm \sqrt{(\Delta_h (1+p^2) - \Delta_\epsilon)^2+4 \Delta_\epsilon \Delta_h}, 0, 2 p \Delta_h)^T$. The spectrum is degenerate at $\epsilon = 0$ where a mean field first order transition takes place, see also Fig.~\ref{SM:fig:DispRel}. We investigate the properties of the phases at $p =0$ (but $h \neq 0$). While the physical properties are similar even when $p \neq 0$, the considered limit allows for a clearer presentation of results.

\paragraph{Case $\epsilon >0$.} $\hat n = \hat e_z$ defines the ground state with $\rho_{\rm MF} = \mu/g_0$. The wave function and observables presented in the section "Characterization of phases" of the main text immediately follow from Eqs.~\eqref{eq:MFWF} and \eqref{eq:MFObs}. Returning to the lab frame, the magnetization follows the external spiral field but the condensate wave function is only weakly rotating:
\begin{equation}
\Psi_{\rm lab} = e^{-i \Theta x S_z} \sqrt{\rho_{\rm MF}} e^{i \vartheta} \left (\hat e_z + \frac{h}{2 g_2 \rho_0} \hat e_y\right ) =  \sqrt{\rho_{\rm MF}} e^{i \vartheta} \left (\hat e_z + \frac{h}{2 g_2 \rho_0} \left (\begin{array}{c}
- \sin(\Theta x) \\ 
\cos(\Theta x) \\ 
0
\end{array} \right )\right ) .
\end{equation}

\paragraph{Case $\epsilon <0$.} In this case $\hat e_y$ is the groundstate for negative $\epsilon$ with $\rho_{\rm MF} = [\mu + \vert \epsilon \vert]/g_0$. Again, the expression presented in the main text follow from Eqs.~\eqref{eq:MFWF} and \eqref{eq:MFObs}.
Returning to the lab frame, the magnetization follows the external spiral field and this time the condensate wave function is also strongly spiral:
\begin{equation}
\Psi_{\rm lab} = e^{-i \Theta x S_z}\sqrt{\rho_{\rm MF}} e^{i \vartheta} \left (\hat e_y - \frac{h}{2 g_2 \rho_0} \hat e_z\right ) = \sqrt{\rho_{\rm MF}} e^{i \vartheta} \left (\left (\begin{array}{c}
- \sin(\Theta x) \\ 
\cos(\Theta x) \\ 
0
\end{array} \right ) - \frac{h}{2 g_2 \rho_0} \hat e_z\right ).
\end{equation}

\section{RG near $(K_s,\Delta_{\epsilon,h}, \lambda_{\epsilon,\Theta,h}) = (\infty,0,0)$}
\label{sec:NLSMRG}

We split $\vartheta \rightarrow \vartheta_s +\vartheta$, $\hat n = O_s O_f \hat e_z$ and expand
$
O_f  = e^{i \not{\alpha}} \simeq 1 + i\not \alpha - \frac{1}{2} [\not \alpha]^2
$  ($\not{\alpha} = \alpha_5 \lambda_5 + \alpha_7 \lambda_7 $).
The fast modes live in the energy-momentum shell defined by 
$
\omega^2/v_s + v_s p^2 \in (\tilde \Lambda^2, \Lambda^2).
$, with $\tilde \Lambda = e^{-b} \Lambda$.
The action splits up as $
S[\vartheta_s +\vartheta,O_sO_f] =S[\vartheta_s ,O_s] + S_2[\vartheta, \not \alpha] + \delta S
$
with  ($\vec p = (\omega, p)$) 
\begin{subequations}
\begin{eqnarray}
S_2[\vartheta, \not \alpha] &=& \frac{1}{2}\int_{\vec p} \frac{K_c}{\pi} \vartheta(-\vec p)\left [ \omega^2 /v_c + v_c p^2\right ]\vartheta(\vec p) + \frac{1}{2}\int_{\vec p} \frac{K_s}{\pi}\sum_{\mu = 5,7} \alpha_{\mu}(-\vec p)\left [ \omega^2 /v_s + v_s p^2\right ]\alpha_{\mu}(\vec p), \\
\delta S &=& \delta S_2 \vert_{ff} +\delta S_2 \vert_{f\partial f} + \sum_{\lambda = \lambda_\epsilon, \lambda_{h}, \lambda_\Theta}\left [\delta S_{1, \lambda} \vert_{ff} +\delta S_{1, \lambda} \vert_{f \partial f} \right ]+ \sum_{\Delta = \Delta_\epsilon, \Delta_h}\delta S_{2, \Delta} \vert_{ff}.
\end{eqnarray}
\end{subequations}
The schematic index `$\vert_{ff}$' (`$\vert_{f\partial f}$') indicates terms with zero (one) derivative acting on a fast field.
Then the corrections due to fast fluctuations take the form
\begin{eqnarray}
S_{\rm eff}[\vartheta_s ,O_s] - S[\vartheta_s ,O_s]  &=& {\llangle \delta S_2  \vert_{ff}\rrangle_{\rm fast} - \frac{1}{2} \llangle \left  [ \delta S_2  \vert_{f\partial f} \right ]^2\rrangle_{\rm fast}} \notag \\
&+&  {\sum_\lambda\llangle \delta S_{1, \lambda} \vert_{ff} \rrangle_{\rm fast} -\sum_\lambda \llangle \left [\delta S_{1, \lambda} \vert_{f\partial f} \delta S_{2}  \vert_{f\partial f} \right ] \rrangle_{\rm fast}} \notag \\
&+&  {\sum_\Delta \llangle \delta S_{2, \Delta} \vert_{ff} \rrangle_{\rm fast}}. \label{eq:Corrections}
\end{eqnarray}
Double angular brackets $\llangle \dots \rrangle_{\rm fast}$ denote connected Wick contraction of fast modes.
{Terms in the first line} are the RG of the NL$\sigma$M, {terms in the second line} yield the scaling dimensions of $\lambda$s, and the third line determines the RG equation of potentials.
The evaluation of Wick contractions over fast fields yields for the usual NL$\sigma$M renormalization
\begin{equation}
{\llangle \delta S_2  \vert_{ff}\rrangle_{\rm fast} - \frac{1}{2} \llangle \left  [ \delta S_2  \vert_{f\partial f} \right ]^2\rrangle_{\rm fast} = \frac{-1}{4\pi} \ln\left (\frac{\Lambda}{\tilde \Lambda}\right ) \int_{\tau, x} \frac{\vert \dot{\hat n}_s\vert ^2}{v_s} + v_s \vert \hat n'_s\vert ^2.}
\end{equation}
Terms with a single derivative lead to the scaling dimensions of $\lambda$s 
\begin{subequations}\begin{eqnarray}
{\llangle \delta S_{1, \lambda_{\epsilon}} \vert_{ff} \rrangle_{\rm fast} - \llangle \left [\delta S_{1, \lambda_{\epsilon}} \vert_{f\partial f} \delta S_{2}  \vert_{f\partial f} \right ] \rrangle_{\rm fast}} &=& - \frac{3}{2 K_s} \ln\left (\frac{\Lambda}{\tilde \Lambda}\right ) \int_{\tau, x} -i \lambda_{\epsilon} \dot \vartheta_s \hat n_s S_z^2 \hat n_s ,\\
{\llangle \delta S_{1, \lambda_h} \vert_{ff} \rrangle_{\rm fast} - \llangle \left [\delta S_{1, \lambda_{h}} \vert_{f\partial f} \delta S_{2}  \vert_{f\partial f} \right ] \rrangle_{\rm fast}} &=& - \frac{1}{2 K_s} \ln\left (\frac{\Lambda}{\tilde \Lambda}\right ) \int_{\tau, x}  \lambda_h \dot{\hat n}_s S_x \hat n_s ,\\
{\llangle \delta S_{1, \lambda_\Theta} \vert_{ff} \rrangle_{\rm fast} - \llangle \left [\delta S_{1, \lambda_\Theta} \vert_{f\partial f}
\delta S_{2}  \vert_{f\partial f} \right ] \rrangle_{\rm fast}} &=& - \frac{1}{2 K_s} \ln\left (\frac{\Lambda}{\tilde \Lambda}\right ) \int_{\tau, x}  i \lambda_\Theta {\hat n}_s' S_z \hat n_s .
\end{eqnarray}\end{subequations}
The scaling dimension of the potentials is obtained from
\begin{subequations}\begin{eqnarray}
{ \llangle \delta S_{0, \Delta_\epsilon} \vert_{f f} \rrangle_{\rm fast} } &=& - \frac{3}{2 K_s}  \ln\left (\frac{\Lambda}{\tilde \Lambda}\right ) \int_{\tau, x}  \Delta_\epsilon\hat n_s S_z^2 \hat n_s ,\\
{ \llangle \delta S_{0, \Delta_h} \vert_{f f} \rrangle_{\rm fast} } &=& - \frac{3}{2 K_s}   \ln\left (\frac{\Lambda}{\tilde \Lambda}\right ) \int_{\tau, x} (- \Delta_h)\hat n_s (S_x+pS_z)^2 \hat n_s .
\end{eqnarray}\end{subequations}

The RG equations near $(K_s,\Delta_{\epsilon,h}, \lambda_{\epsilon,\Theta,h}) = (\infty,0,0)$, which we discussed in the beginning of section "Characterization of phase transitions" of the main text, immediately follow immediately follow from these expressions.

\section{RG in the easy plane: Derivation of Equations (4) of the main text.}

The inclusion of phase slips in the easy plan action leads to
\begin{equation}
S = \int_{\tau, x} \frac{K_c}{2\pi} \left [\frac{\dot \vartheta^2}{v_c} +v_c {\vartheta'}^2\right ] + \frac{K_c}{2\pi} \left [\frac{\dot \phi^2}{v_c} +v_c {\phi'}^2\right ] - \frac{\Delta_\epsilon - i \lambda_\epsilon \dot \vartheta}{2} \frac{1}{1+p^2}\cos(2\phi) + \frac{y}{2} \cos(\tilde \phi) - \frac{i}{2\pi} \vec \nabla \phi \wedge \vec \nabla \tilde \phi.
\end{equation}
We have introduced the dual field $\tilde \phi$ in this equation. In the following and in the main text, the factor of $(1+p^2)^{-1}$ is reabsorbed into $\Delta_\epsilon, \lambda_\epsilon$. Again, we split fields into fast and slow, $\phi_s + \phi_f$, analogously for $\vartheta$ and $\tilde \phi$, and expand the cosines to second order in fast fields. Subsequently, fast fields are integrated out. This yields the effective action
$
S_{\rm eff}= S_s + \llangle \delta S \rrangle_{\rm fast} + \llangle  \delta \tilde S \rrangle_{\rm fast} - \frac{1}{2} \llangle (\delta S + \delta \tilde S)^2 \rrangle_{\rm fast},
$
and we introduced 
\begin{subequations}\begin{eqnarray}
\delta S &\simeq& \int_{\tau, x} -\frac{\Delta_\epsilon - i \lambda_\epsilon \dot \vartheta_s}{2}[-\cos(2\phi_s) 2 \phi_f^2 - \sin(2 \phi_s)2 \phi_f] + \int_{\tau, x}  i \lambda_\epsilon \frac{\dot \vartheta_f}{2}[\cos(2\phi_s) (1 - 2 \phi_f^2) - \sin(2 \phi_s)2 \phi_f], \label{eq:deltaS} \\
\delta \tilde S &\simeq& \int_{\tau, x} \frac{y}{2}[-\cos(\tilde \phi_s) \tilde \phi_f^2/2 - \sin(\tilde \phi_s) \tilde \phi_f].
\end{eqnarray}\end{subequations}
We use the same regularization scheme as above. Then $\llangle \delta S \rrangle_{\rm fast}$ and $\llangle  \delta \tilde S \rrangle_{\rm fast}$ readily provide the rescaling of $\Delta_\epsilon, \lambda_\epsilon, y$
\begin{subequations}\begin{eqnarray}
\Delta_\epsilon &\rightarrow& \Delta_\epsilon(1 - 2 \langle \phi_f^2 \rangle )  = \Delta_\epsilon ( 1- K_s^{-1} \ln(\Lambda/\tilde \Lambda)) \\
\lambda_\epsilon &\rightarrow& \lambda_\epsilon(1 - 2 \langle \phi_f^2 \rangle )  = \lambda_\epsilon ( 1- K_s^{-1} \ln(\Lambda/\tilde \Lambda)) \\
y &\rightarrow& y(1 -  \langle\tilde \phi_f^2 \rangle/2 )  = y ( 1- K_s \ln(\Lambda/\tilde \Lambda)).
\end{eqnarray}\end{subequations}
To obtain the renormalization corrections of the kinetic terms we use that, for a general slow function $f_s(\v r)$,
\begin{subequations}\begin{eqnarray}
\llangle \int_{\v r, \v r'} f_s(\v r) \cos(\tilde \phi(\v r)) f_s(\v r') \cos(\tilde \phi(\v r')) \rrangle_{\rm fast} &\simeq& -\frac{\pi}{2 \Lambda^4} \ln\left (\frac{\Lambda}{\tilde \Lambda}\right ) \int_{\v R} f_s(\v R)^2 [(\nabla \tilde \phi_s)^2C_3(K_s) - 4 \Lambda^2 C_1(K_s)] , \\
\llangle \int_{\v r, \v r'} f_s(\v r) \cos(2 \phi(\v r)) f_s(\v r')\cos(2 \phi(\v r')) \rrangle_{\rm fast} &\simeq& -\frac{\pi}{2 \Lambda^4} \ln\left (\frac{\Lambda}{\tilde \Lambda}\right ) \int_{\v R} f_s(\v R)^2 [(\nabla \phi_s)^2C_3(1/K_s) - 4 \Lambda^2 C_1(1/K_s)],
\end{eqnarray}\end{subequations}
and we introduced (see e.g. T. Giamarchi, "Quantum Physics in One Dimension", Clarendon Press, 2004)
\begin{equation}
C_n(z) = z \int_0^\infty dx x^n e^{- 2 z F_1(x)} J_0(x)  \text{ with } F_1(x) = \int_0^1 \frac{dq}{q} (1 - J_0(qx)).
\end{equation}
We consider the RG equations only to leading (i.e. zeroth order) in $1/K_c \ll 1$, so that the second term in Eq.~\eqref{eq:deltaS} may be omitted. As usual, the renormalization of kinetic terms by means of the cosine terms involves non-universal, weakly $K_s$-dependent prefactors. We absorb those into a redefinition of $\lambda_\epsilon, \Delta_\epsilon, y$ and omit non-universal corrections to the RG equations stemming from $d[\ln C_n(K)]/ db$ etc. Then we readily obtain Eqs. (4) of the main text.

\section{Fermionization and derivation of RG equations (6) of the main text.}

The fermionization dictionary of Ref.~[52] demonstrates the Ising nature of the finite epsilon transition. At $K_s = 1$ and $\lambda_\epsilon = 0$ we represent the spin sector by means of a spinless Dirac fermion $\psi$
\begin{equation}
S_{\rm spin}  = \frac{1}{2} \int_{\tau, x} \left (\psi^\dagger, \psi^T \sigma_x\right ) \left (\begin{array}{cc}
\partial_\tau + v_s \hat p \sigma_z + M_\epsilon \sigma_x& -M_v \sigma_z \\ 
-M_v \sigma_z & \partial_\tau -[ \hat p \sigma_z +M_\epsilon \sigma_x]
\end{array} \right ) \left (\begin{array}{c}
\psi \\ 
\sigma_x (\psi^\dagger)^T
\end{array} \right ) . 
\end{equation}
We now define 
\begin{equation}
\eta = \left (\begin{array}{cccc}
1 & 0 & 0 & 0 \\ 
0 & -i & 0 & 0 \\ 
0 & 0 & -i & 0 \\ 
0 & 0 & 0 & 1
\end{array} \right ) \frac{1}{\sqrt{2}}\left (\begin{array}{cccc}
1 & 0 &0 & 1 \\ 
0 & -1 & 1 & 0 \\ 
-1 & 0 & 0 & 1 \\ 
0 & 1 & 1 & 0
\end{array} \right ) \left (\begin{array}{c}
\psi \\ 
\sigma_x (\psi^\dagger)^T
\end{array} \right )
\end{equation}
and restore $\lambda_\epsilon$ by $M_\epsilon \rightarrow M_\epsilon - i \lambda \dot \vartheta$ to obtain Eq.~(5) of the main text.

For the RG treatment of Eq.~(5) of the main text, we denote $M_\epsilon + M_v \equiv m$, assume $\vert m \vert \ll  M_v- M_\epsilon$ and discard the other Majorana modes. To make connection to Refs.~[44-46] we perform a Lorentz-Boost $(v_s \tau, x) \rightarrow (x,- v_s \tau)$ and subsequently rotate $\eta_{R,L} \rightarrow e^{\pi i \pi/2} \eta_{R,L}$. This leads to 
\begin{equation}
S = \frac{1}{2} \int_{\tau, x} \frac{1}{\tilde K\pi} \left [\frac{\dot \vartheta^2}{\tilde v} + \vartheta'^2 \tilde v\right ] + \eta^T \left [\partial_\tau + v_s \hat p \sigma _z + (m +2 i \tilde \lambda  \vartheta')\sigma_y \right ] \eta. 
\end{equation}
In this equation, $\eta$ is a real two-spinor, and $\tilde K = K_c^{-1}, \tilde \lambda = v_s \lambda/2, \tilde v = v_s^2/v_c$. Note that the imaginary $i$ is not present in the model studied in Refs.~[44-46].

Again, the Wilsonian RG is performed by splitting $\vartheta$ and $\eta$ in slow and fast fields. The interaction terms entering $\delta S = \delta S_1 + \delta S_2$ are
$
\delta S_1 = i \tilde \lambda \int_{\tau, x} 2 \vartheta_f' \eta_f^T \sigma_y \eta_s ,
\delta S_2 = i \tilde \lambda \int_{\tau, x} \vartheta_s' \eta_f^T \sigma_y \eta_f.
$
The effective action obtained at one-loop approximation
\begin{equation}
S_{\rm eff} = S_{\rm slow} - \frac{1}{2} \llangle \delta S^2 \rrangle_{\rm fast} + \frac{1}{3!} \llangle \delta S^3 \rrangle_{\rm fast}
\end{equation}
contains a ``bubble'' renormalization the fermionic propagators ($-\llangle \delta S_1^2 \rrangle_{\rm fast}/2$), a ``bubble'' renormalization the bosonic propagators ($-\llangle \delta S_2^2 \rrangle_{\rm fast}/2$), and a triangle diagram renormalizing $\tilde \lambda$ ($\llangle \delta S_1^2 \delta S_2 \rrangle_{\rm fast}/2$). After performing Wick contraction's we obtain
\begin{equation}
S_{\rm eff} -S_{\rm slow}=\frac{1}{2} \int_{\tau, x} \Bigg \lbrace\frac{\vartheta'^2 \tilde v }{\tilde K\pi} \left ( \frac{4 \tilde \lambda^2 \pi \tilde K I_3}{\tilde v}\right ) + \eta^T \left [\partial_\tau ( 4 \tilde \lambda^2 \pi \tilde K I_2) + v_s \hat p \sigma _z (4 \tilde \lambda^2 \pi \tilde K I_2)+ (m +2 i \tilde \lambda  \vartheta')\sigma_y ( 4 \tilde \lambda^2 \pi \tilde K I_1)\right ] \eta \Bigg \rbrace. 
\end{equation}
We introduced the logarithmic integrals 
\begin{equation}
I_1 = \frac{\ln(\Lambda/\tilde \Lambda)}{2\pi v_s^2}\frac{1}{1+\tilde v/v_s}, \; 
I_2 =  \frac{\ln(\Lambda/\tilde \Lambda)}{2\pi v_s^2}\frac{1}{(1+\tilde v/v_s)^2}, \;
I_3 =  \frac{\ln(\Lambda/\tilde \Lambda)}{2\pi v_s}.
\end{equation}
Under the identification $\tilde \lambda \rightarrow - i \lambda/2, \tilde K \rightarrow K_\rho, v_s \rightarrow u, \tilde v \rightarrow v$ these expressions are consistent with the RG equations of Ref.~[45]. We now restore the notation of the main text
$G^2 = \frac{\lambda^2}{K_c} = 4\frac{\tilde \lambda^2 \tilde K}{v_s^2},
u = \frac{v_c}{v_s}  = \frac{v_s}{\tilde v} ,
\bar v =  \sqrt{v_c v_s} = v_s \sqrt{\frac{v_s}{\tilde v}} ,
K_c = 1/\tilde K,
$
then Eqs.~(6) of the main text immediately follow.

\section{Critical behavior of observables and comparison to experiment}

In this section we summarize the critical behavior of nematic observables $N_{zz}- N_{yy} = \cos(2\phi)$ and for the  order parameter $\hat n_{z} = \cos(\phi)$. Since our theory demonstrates the independence of critical properties on small $p$, we concentrate on the case $p = 0$. Later in this section we discuss our theory in light of realistic experimental setups. 

\textbf{Phase transition at $K_s \geq 2$.}
	 We first consider the case when $K_s$ is strictly larger than 2, in this regime, vortices are irrelevant in the RG sense. The integration of the RG equation $\partial \Delta / \partial b = (2-1/K_s) \Delta$ in the approximation of $K_s (\Lambda) \simeq const.$ leads to 
$
	\Delta(\Lambda) = \Delta_0 (\Lambda_0/\Lambda)^{{2 -1/K_s}}
$
	Note that we here used the notation $\Delta$ for the running coupling constant containing the dimensionful rescaling of $\Delta_\epsilon$, at bare level $\Delta_0 = \Delta (\Lambda_0) \sim {2 \pi \Delta_\epsilon}/({K_s \Lambda_0^2})$. Therefore, the RG stops at the scale where $\Delta(\Lambda) = 1$ i.e. at 
$
	\Lambda_\epsilon = \Lambda_0 \vert\Delta_0\vert^{\frac{1}{2-1/K_s}}.
$
	Beyond this scale, bosonic fluctuations can be integrated at the level of the Gaussian approximation near $\phi = 0$ ($\phi = \pi/2$) in the case $\epsilon >0$ ($\epsilon <0$). This leads to 
	\begin{equation}
		\langle \Delta N  \rangle = \text{sign}(\epsilon)\left (\frac{\vert\Delta(\Lambda_\epsilon)\vert\left (\frac{\Lambda_\epsilon}{\Lambda_0}\right )^2} {1+ \vert\Delta(\Lambda_\epsilon)\vert \left (\frac{\Lambda_\epsilon}{\Lambda_0}\right )^2}\right )^{\frac{1}{2K_s}}, \;  \langle n_z  \rangle = \text{sign}(\epsilon)\left (\frac{\vert\Delta(\Lambda_\epsilon)\vert\left (\frac{\Lambda_\epsilon}{\Lambda_0}\right )^2} {1+ \vert\Delta(\Lambda_\epsilon)\vert \left (\frac{\Lambda_\epsilon}{\Lambda_0}\right )^2}\right )^{\frac{1}{8K_s}}.
	\end{equation}
	Using the given definition of $\Lambda_\epsilon$ this readily implies the critical exponents presented in the caption of Fig.~1.
	We now discuss the case $K_s = 2$: In this case, $y$ is marginal, but $\Delta_\epsilon$ is relevant. Therefore,  any infinitesimal $\epsilon$ dominates the scaling and our result is applicable. At $\epsilon = 0$, $\langle \Delta N  \rangle =0$ by symmetry. 
	
\textbf{Ising transition for $K_s <2$}. We begin by considering $\Delta N$, and for the moment $\lambda_\epsilon = 0$. We remark that $N_{zz}$ may be obtained from the fermionic theory by differentiating the partition function with respect to $\Delta_\epsilon$. This yields $\Delta N \sim \langle \tr \sigma_y \kappa_z \eta \eta^T \rangle \sim M_\epsilon$ up to logarithmic corrections, i.e. linear behavior. To obtain a crossover formula one may introduce two replicas of the Majorana fermions $\eta_1, \eta_2$ and combine them into a complex fermion $\psi = (\eta_1 + i \eta_2)/\sqrt{2}$. This fermion may then be bosonized so that the theory is determined by the effective Lagrangian 
	\begin{equation}
	\mathcal L = \sum_{\kappa = \pm}\left [\frac{1}{2\pi} ([\dot \varphi_\kappa]^2/v_s + [\varphi_\kappa']^2) + (\Delta_\epsilon + \kappa y) \sin^2 (\varphi_\kappa)\right ].
	\end{equation}
	Differentiating the replicated partition function with respect to $\Delta_\epsilon$ yields the nematic order parameters, so that
	\begin{equation}
	\langle \Delta N \rangle = \frac{1}{2} \sum_\kappa \langle \sin^2 \varphi_\kappa \rangle_\varphi = \frac{1}{2} \sum_\kappa \text{sign}(\Delta_\epsilon + \kappa y) \left (\frac{\frac{2\pi}{\Lambda^2}\vert \Delta_\epsilon + \kappa y \vert}{1+\frac{2\pi}{\Lambda^2}\vert \Delta_\epsilon + \kappa y \vert}\right )^{1/2}.
	\end{equation}
	Note that $\Delta_\epsilon \rightarrow \Delta_\epsilon - i \lambda_\epsilon \dot \vartheta $ does not change the result because $\lambda_\epsilon$ drops out at Gaussian order. Furthermore, employing fermionic RG prior to rebosonization, it follows that $\lambda_\epsilon$ is irrelevant and $\Delta_\epsilon + \kappa y$ has scaling dimension of unity (up to corrections which are slower than powerlaw [44]). Therefore,  
	\begin{equation}
	\langle \Delta N \rangle \simeq \frac{1}{2} \sum_\kappa \text{sign}(\Delta_\epsilon + \kappa y) \left (\frac{\left (\frac{2\pi}{\Lambda_0}\vert \Delta_\epsilon + \kappa y \vert\right )^2}{1+\left (\frac{2\pi}{\Lambda_0}\vert \Delta_\epsilon + \kappa y \vert\right )^2}\right )^{1/2}.
	\end{equation}
	
	To determine the power law of the order parameter field we write the low energy theory $(0,\sin \phi, \cos \phi) \rightarrow \vec \Phi = (\Phi_y, \Phi_z)^T$ as a two component $\Phi^4$ theory with a symmetry breaking field
	\begin{equation}
	\mathcal L = \frac{K_s}{2\pi}([\dot{\vec \Phi}]^2/v_s + [\vec \Phi']^2 v_s) - \alpha \vec \Phi^2 + \frac{\beta}{2} \vec \Phi^4 + \Delta_\epsilon \Phi_y^2. \label{eq:Phi4}
	\end{equation}
	In our case, $\alpha = \beta \propto \Lambda^2$, are the highest scales in the and impose $\vec \Phi^2 =1$ near $\epsilon =0$. Note that phase slips are implicitly included in the theory~\eqref{eq:Phi4}, as they correspond to Abrikosov solutions of the Ginzburg-Landau equations. 
	
	The symmetry breaking term projects Eq.~\eqref{eq:Phi4} to the easy axes and integration of $\Phi_y$ assuming $\Delta_\epsilon - \alpha>0$ leads
	\begin{equation}
	\mathcal L = \frac{K_s}{2\pi}(\dot{\Phi_z}^2/v_s + {\Phi'_z}^2 v_s) + (\beta \ln\left (1+\frac{\Lambda^2}{\Delta_\epsilon - \alpha}\right )-\alpha)  \Phi_z^2 + \frac{\beta}{2}  \Phi_z^4. \label{eq:Phi4}
	\end{equation}	
	The mass term changes sign from positive to negative at some $\Delta_{\epsilon_I}/\alpha \sim \mathcal O(1)$ (but clearly $\Delta_{\epsilon_I}>\alpha$).
	It is known that Eq.~\eqref{eq:Phi4} describes an Ising transition in 1+1D. We thus identify $\langle n_z \rangle = \langle \Phi_z \rangle$ with the Ising order parameter at the transition and therefore conclude that near the Ising transition
	\begin{equation}
	\langle n_z \rangle \sim H(\Delta_\epsilon - y)  \vert \Delta_\epsilon - y \vert^\frac{1}{8}.
	\end{equation}
	This concludes the derivation of critical exponents reported in the caption of Fig.~1.

\textbf{Experimental realization, finite size and finite temperature.} We here include an estimate of the length scales and Luttinger parameters for  realistic experiments using parameters given in Tab. II and Eqs.~(11),~(12) and (17) of Ref.~[5]. In addition to the length scales defined in the main text we consider the transversal radius, $l_\perp = \sqrt{\hbar/(m \omega_\perp)}$, the longitudinal system size $l_{\rm trap} = \sqrt{2\mu/(m\omega_\Vert^2)}$ (with typical trapping frequencies $\omega_\Vert \sim 10$ Hz and $\omega_\perp \sim 100$ kHz) as well as the thermal length $l_T = v_s/T$ (with typical temperature $T \sim 1$ nK). These length scales, as well as bare Luttinger parameters are plotted in Figs.~\ref{SM:fig:Lengthscales} (a) and (b), and are rather illuminating. The first observation is that the length scale set by the trap is the longest length scale in the problem. For the Luttinger Liquid in the charge sector to remain a valid description, the system needs to be at a temperature such that $v_c/T > \xi_c$. Similarly, for the low energy description in the spin sector, and the proposed quantum critical properties, to be experimentally accessible requires temperatures $v_s/T > \xi_s$, which as shown in Fig.~\ref{SM:fig:Lengthscales} a) requires on the order of $N \sim 100$ bosons per tube.

We now discuss details on the finite temperatures $T$ and of a confining harmonic trap, the latter being incorporated in the chemical potential $\mu \rightarrow \mu(x) = \mu (1- x^2/l_{\rm trap}^2)$ with $x \in (-l_{\rm trap},l_{\rm trap})$. Our theory assumes the trap frequency $\omega_\Vert$ to be weak as compared to interactions, i.e. $l_{\rm trap} \gg K_{c}^2 \xi_{c}$, keeping the hierarchy of length scales, Fig.~2, otherwise intact. In this case the mean field equation may be solved locally in the semiclassical approximation, leading to a Thomas-Fermi BEC with a local superfluid density profile $\rho_0 \bar \rho (x)$ and $\bar \rho(x) = (1- x^2/l_{\rm trap}^2)$. The parameters entering Eq.~(2) of the main text are replaced by $\Delta_\epsilon \rightarrow \Delta_\epsilon \bar \rho(x) , \Delta_h \rightarrow \Delta_h ,
\lambda_{\epsilon,h} \rightarrow \lambda_{\epsilon,h} , \lambda_\theta \rightarrow \lambda_\theta\bar \rho(x)  ,
K_{c,s} \rightarrow K_{c,s} \sqrt{\bar \rho(x)} , v_{c,s} \rightarrow v_{c,s} \sqrt{\bar \rho(x)} $.
This effect becomes apparent in the Green's functions (for simplicity, here $v_s = v_c = 1$) defined by
\begin{equation}
-\frac{1}{\pi} \left (\partial_\tau^2 G(\tau; x,x') + \partial_x [\bar \rho(x) \partial_x G(\tau; x,x')]\right ) + \frac{m^2}{\pi}  G(\tau; x,x') = \delta(\tau) \Delta(x - x').
\end{equation}
The solution may be in general represented using the mode expansion in terms of Legendre Polynomials
\begin{equation}
G(\tau; x,x') = \frac{\pi T}{l_{\rm trap}} \sum_{k = -\infty}^\infty \sum_{n = 1}^\infty \frac{e^{i 2\pi T k \tau} }{(2\pi T k)^2 + n(n+1)/l_{\rm trap}^2 + m^2} \frac{2n+1}{2} P_n(x/l_{\rm trap}) P_n(x'/l_{\rm trap}). \label{SM:eq:Correl}
\end{equation}
The logarithmic Green's functions are thus exponentially cut-off at the smallest of the three length scales $l_{\rm trap}$, $1/T$ or $1/m$.
This expression readily provides the physical correlation functions of interest. For example, nematic fluctuations in the easy plane theory near $\epsilon = 0$ (in this case $m = \pi \sqrt{2 \vert \Delta_\epsilon \vert}$) are characterized by
\begin{equation}
\langle (N_{zz}-N_{yy})_{\tau,x} (N_{zz}-N_{yy})_{0,x'} \rangle \sim e^{-2 \langle [\phi (\tau,x) - \phi(0, x')]^2 \rangle} \sim e^{4 G (\tau; x, x')/K}. \label{eq:NCorrel}
\end{equation}

In Fig.~(1) of the main text we present the critical behavior of the order parameter at the transition, which, as explained in the previous section, is obtained from the integration of RG equations. This procedure captures the leading asymptotics and the scale of the cut-off while, contrary to Eq.~\eqref{eq:NCorrel}, details on the precise functional behavior of $G(\tau;x, x')$ are irrelevant. We return to the example of $K_s >2$, and following Fig.~\eqref{SM:fig:Lengthscales} a), use $l_T \ll l_{\rm trap}$. The powerlaw behavior remains unchanged for $\xi_\epsilon = \sqrt{v_s/\Delta_\epsilon} < l_T$, but in the small window $l_T < \xi_\epsilon$, the correlator Eq.~\eqref{SM:eq:Correl} becomes essentially one dimensional, therefore $\langle \Delta N \rangle$ vanishes and as a consequence the quantum critical behavior is distorted, see Fig.~\eqref{SM:fig:Lengthscales} c). 

\begin{figure}
  \begin{center}
    \includegraphics[width=0.9\textwidth]{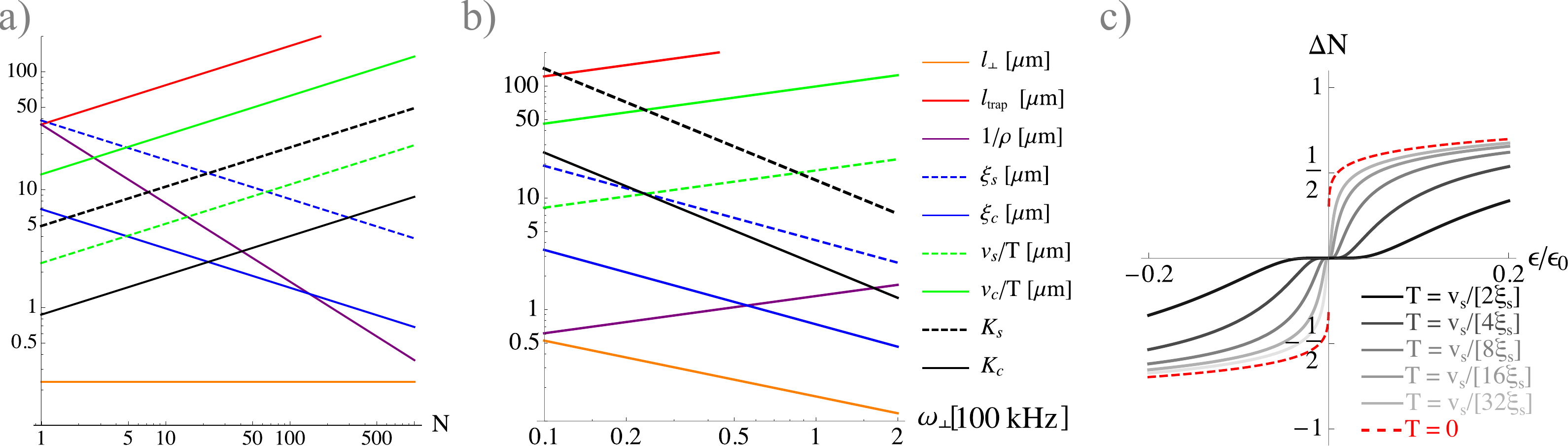} 
  \end{center}
  \caption{Comparison to realistic experimental setups. Panel a): Estimate of length scales and Luttinger parameters as a function of the number $N$ of bosons in a single 1D tube. For this plot, we chose $\omega_\Vert = 10$ Hz, $\omega_\perp = 50$ kHz, $T = 0.2$ nK and parameters as for $^{23}$Na~[5]. Based on the given length scales, quantum criticality should be observable for $N \gtrsim 100$. 
  Panel b): Estimate of length scales and Luttinger parameters as a function of the trapping frequency $\omega_{\perp}$ perpendicular to the tubes of atoms. The range of length scales implies that it is advantageous to use an $\omega_{\perp}\approx 100$ kHz trapping frequency to reduce the rounding of the quantum critical properties.
  Panel c): Rounding of the transition for various temperatures, all for $K_s = 5$. The intermediate flat region is a manifestation of classical disorder in 1D at finite temperature.}
  \label{SM:fig:Lengthscales}
\end{figure}

\newpage
\end{widetext}

\end{document}